\documentclass[runningheads]{llncs}
\usepackage[T1]{fontenc}
\usepackage{graphicx}
\begin{document}
%
%\title{Biologically-inspired parameter-free channel attention for cardiac MRI reconstruction}

\titlerunning{Parameter-Free Biologically-Inspired Channel Attention}
\title{Parameter-Free Bio-Inspired Channel Attention for Enhanced Cardiac MRI Reconstruction}
% If the paper title is too long for the running head, you can set
% an abbreviated paper title here
%
\author{Anam Hashmi\inst{1}\orcidID{0009-0007-0887-246X} \and
Julia Dietlmeier\inst{2}\orcidID{0000-0001-9980-0910} \and
Kathleen M. Curran\inst{3}\orcidID{0000-0003-0095-9337} \and
\\ Noel E. O'Connor\inst{2}\orcidID{0000-0002-4033-9135}}
\authorrunning{Anam Hashmi et al.}
% First names are abbreviated in the running head.
% If there are more than two authors, 'et al.' is used.
%
\institute{ML-Labs, Dublin City University, Ireland \\ 
\email{anam.hashmi2@mail.dcu.ie} \and
Insight SFI Research Centre for Data Analytics, Dublin City University, Ireland \\
\email{\{julia.dietlmeier,noel.oconnor\}@insight-centre.org}\\ \and
School of Medicine, University College Dublin, Ireland\\
\email{kathleen.curran@ucd.ie}}
\maketitle              % typeset the header of the contribution
\begin{abstract}
Attention is a fundamental component of the human visual recognition system. The inclusion of attention in a convolutional neural network amplifies relevant visual features and suppresses the less important ones. Integrating attention mechanisms into convolutional neural networks enhances model performance and interpretability. Spatial and channel attention mechanisms have shown significant advantages across many downstream tasks in medical imaging. While existing attention modules have proven to be effective, their design often lacks a robust theoretical underpinning. In this study, we address this gap by proposing a non-linear attention architecture for cardiac MRI reconstruction and hypothesize that insights from ecological principles can guide the development of effective and efficient attention mechanisms. Specifically, we investigate a non-linear ecological difference equation that describes single-species population growth to devise a parameter-free attention module surpassing current state-of-the-art parameter-free methods.

%In this work, we aim to bridge this gap and first target nonlinear attention design in cardiac MRI reconstruction task. Second, we hypothesize that an accurate and efficient attention mechanism can be designed using a wealth of knowledge from the biological domain. Specifically, we investigate a nonlinear ecological difference equation that describes single-species population growth and design a parameter-free attention layer that outperforms the state-of-the-art. 

\end{abstract}
\section{Background and Motivation}
Attention mechanisms play a pivotal role in enhancing the efficacy of convolutional neural networks (CNNs) by focusing on relevant visual features. %Attention may come in different architectures but it serves the same objective to improve the relevant visual features being processed by a convolutional neural network (CNN). 
Attention mechanisms have been successfully integrated as plug-and-play modules into CNNs for various applications such as medical image classification \cite{classification_2022}, segmentation \cite{segmentation_2021}, explainability \cite{interpretability_2022} and most recently Cardiac Magnetic Resonance (CMR) reconstruction \cite{Hashmi_attention_arxiv,GNA_UNET}. % applications can report successful integration of attention mechanisms which are usually designed as plug-in modules that can be easily added to a CNN. 
Several prominent architectures include the squeeze-and-excitation networks (SE) \cite{SE}, linear context transform block (LCT) \cite{LCT}, and convolutional block attention module (CBAM) \cite{CBAM}. These and other state-of-the-art attention architectures %have demonstrated success in different applications, each tailored to specific tasks and requirements. and other state-of-the-art attention architectures 
\cite{AB,SimAM,ECA,GaussianCT} have demonstrated success in different applications, each tailored to specific tasks and requirements. While these attention architectures, along with others, exhibit promising results and some are even parameter-free \cite{SimAM,GaussianCT}, their design often lacks a solid theoretical foundation. Thus, despite their efficacy, there remains a need for a more rigorous theoretical basis to underpin their design choices.
%However, although they all are well motivated they lack solid theoretical foundation to support their design choice.  

%\section{Biologically inspired attention}
\section{Biologically-Inspired Attention }
In this work, we draw inspiration from non-linear ecological difference equations used in population biology \cite{May_ecology_1975}. These mathematical equations describe the dynamical system of growth of a given population depending on several environmental factors. Specifically, the following Equation (1) \cite{May_ecology_1975} (Fig.1 shows implementation in the blue box) describes single-species population growth:
\begin{equation}
N_{t+1}=\lambda [1+\alpha N_t]^{-b}N_t
\end{equation}
\begin{equation}
N_{t+1}=\lambda [1+\alpha N_{t1}]^{-b}N_{t2}
\end{equation}
%\begin{equation}
%N_{t+1}=\frac{\lambda N_t}{1+(\alpha N_t)^b}
%\end{equation}
\begin{figure}
\centering
\vspace{-25pt}
\includegraphics[width=8.5cm]{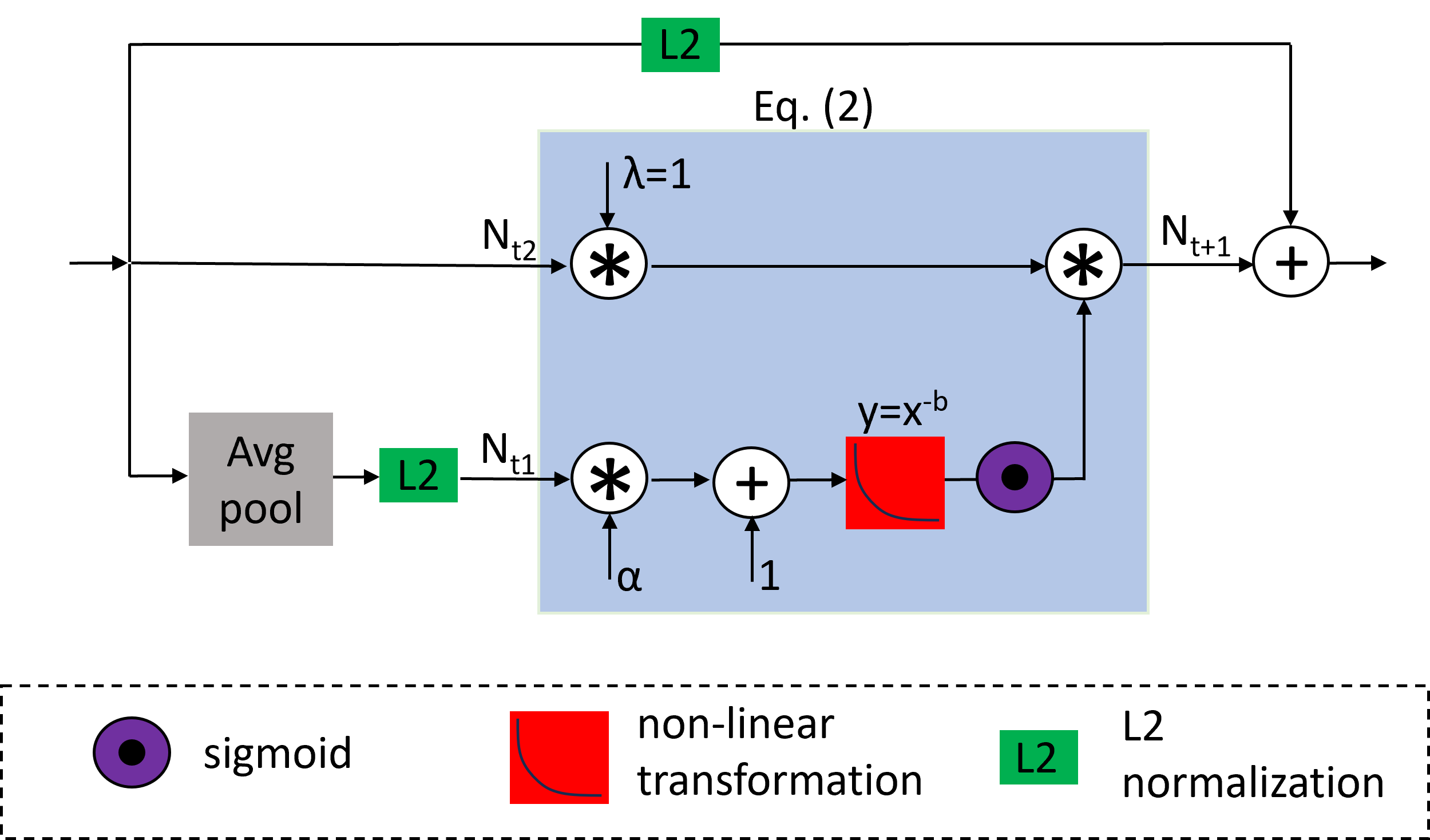}
\caption{A schematic of the proposed biologically-inspired attention block. Our systematic design is based on modelling a non-linear ecological difference equation \cite{May_ecology_1975} in addition to average pooling, sigmoid activation function and L2-normalization layers. This follows the original Eq. (1) but with two separate branches $N_{t1}$ and $N_{t2}$ as defined in Eq. (2). Note that $\alpha=2$ and $b=2$. Optimizing the hyperparameter values could yield better results, but is beyond the scope of this work.}%Our attention layer is entirely parameter-free.}%
\label{fig1}
\end{figure}
%\section{Architecture for cardiac MRI reconstruction}
%\label{sec:Architecture}
\vspace{-10pt}

The logic behind using non-linear ecological equations to model attention draws upon the principles of dynamic adaptation observed in natural systems. Biological populations dynamically adapt to environmental changes, emphasizing relevant factors and suppressing less important ones, similar to the objective of attention mechanisms in neural networks. Moreover, ecological systems, like population dynamics, often show complex non-linear behavior and integrating these dynamics allows to capture the real-world complexity more accurately. %Ecological systems, such as population dynamics, often exhibit complex non-linear behaviour and integrating insights from non-linear difference equations allows for the creation of attention mechanisms that emulate the adaptive behaviour observed in natural systems.
Thus, we hypothesize that non-linear ecological equations provide a framework for capturing the complex dynamic interactions between different visual features.
%For the reconstruction task, we utilize a CMR reconstruction network from \cite{Hashmi_attention_arxiv}. %GNA-UNET \cite{GNA_UNET,SOTA_CMR}. We build a 2D GNA-UNET with four downsampling/upsampling stages and 32 feature maps in the initial downsampling stage. To regularize our model, we add \textbf{Dropout} layers with Dropout probability $p=0.25$ in each of four encoder blocks. All Group Normalization (GN) layers have hyperparameter $ng=8$. All attention blocks (\textbf{Att}) reviewed and proposed in this paper are integrated into the convolutional block of the GNA-UNET model as follows: \textbf{conv2d} $\rightarrow$ \textbf{Att} $\rightarrow$ \textbf{GN} $\rightarrow$ \textbf{conv2d} $\rightarrow$ \textbf{Att} $\rightarrow$ \textbf{GN} $\rightarrow$ \textbf{ReLU}. 
%We term the configuration without attention layers as our \textit{baseline} model. 

\section{Methodology and Experiments}

We experiment with the recently released CMRxRecon dataset \cite{CMRxRecon_arxiv} and obtain quantitative validation metrics such as $PSNR$, $MSE$ and $SSIM$ \cite{metrics}. For the reconstruction task, we utilize a CMR reconstruction network from \cite{Hashmi_attention_arxiv} and term the configuration without attention layers as the \textit{baseline} model. %We perform experiments on the training set of CMRxRecon using the long-axis fully sampled (Ground Truth) and undersampled single-coil data with the acceleration factor AccF of $\times 10$ as these data contain the strongest aliasing artifacts. 
Our methodology follows \cite{GNA_UNET,Hashmi_attention_arxiv}. The preprocessing and training details mirror those in \cite{Hashmi_attention_arxiv}. % but we use $256 \times 256$ pixels input resolution.
Our computing pipeline was implemented using Python 3.11.5 and PyTorch 2.1.1.% framework. %Training details are identical to those in \cite{Hashmi_attention_arxiv}. %We train the U-Net using an AdamW optimizer with a learning rate of $lr=0.001$. % and the Exponential Learning Rate Scheduler with $\gamma=0.95$. 
%To ensure the reproducibility, all seeds are set to 42. Batch size is 2 and all models are trained for 300 epochs to ensure convergence. We use MSE loss as the main optimization objective and do not apply data augmentation. %All images were resized to the $256 \times 256$ image resolution.
%\subsection{Experimental Results}
%Table~\ref{tab1} gives a summary of quantitative results using objective metrics \cite{metrics}.% and Fig.~2 shows qualitative results (patient xxx, short-axis, AccF=xxx, single-coil). %Our data preparation and the training-validation split follows \cite{GNA_UNET}. %We save the best checkpoint by monitoring the training loss (due to its high fluctuations).
\vspace{-10pt}
\begin{table}
\centering
\caption{Comparison of quantitative results. %For fairness, the comparison is provided with the most efficient existing methods. %
Wilcoxon signed-rank test was conducted to compare the SSIM scores of our method and the competing method SimAM. The results indicated a significant difference between the two methods (p-value < 0.001), demonstrating that our method significantly outperforms SimAM.%Computational overhead is given in terms of model parameters. %Inference time is given per patient (LAX, Acc04, P001) and is in seconds.% 
}
\label{tab1}
\begin{tabular}{|l|c|c|c|c|c|}
\hline
Attention &   Computational overhead  & PSNR $\uparrow$ & MSE $\downarrow$ & SSIM $\uparrow$\\
\hline
\hline
baseline &  0  & 36.2068 & 0.0002878 & 0.9245 \\
\hline
+ SE \cite{SE} & 119,936  & 37.2863 & 0.0002355 & 0.9429\\
+ LCT \cite{LCT}& 6,848 & 36.7562 & 0.0002618  & 0.9450\\
+ AB \cite{AB}& 3,424  & 34.3633 & 0.0004299 & 0.9357  \\
+ ECA \cite{ECA} & 90  & 37.9982 & 0.0002262 & 0.9527  \\
\hline
\hline
+ SimAM \cite{SimAM} & 0  & 37.0492 & 0.0002583 & 0.9443  \\
+ GCT \cite{GaussianCT} & 0  & 36.5874 & 0.0002695 & 0.9408  \\
\hline
+ Proposed & 0  & \textbf{37.7724} & \textbf{0.0002231} & \textbf{0.9496} \\
%Proposed B & 0 & 0.7137 & 38.7879 & 0.0001998 & 0.9577\\
\hline
\end{tabular}
\end{table}
\vspace{-10pt}

%As can be seen from Table~\ref{tab1}, the proposed biological attention outperformed the other parameter-free state-of-the-art channel attention blocks reviewed. We have also verified that making hyperparameters $\alpha$, $\lambda$ and $b$ learnable improves the overall performance but increases the computational overhead. Tuning these hyperparameters %or changing the activation function %
%may also impact the performance.

\begin{figure}[h]
\centering
\vspace{-25pt}
\includegraphics[width=10cm]{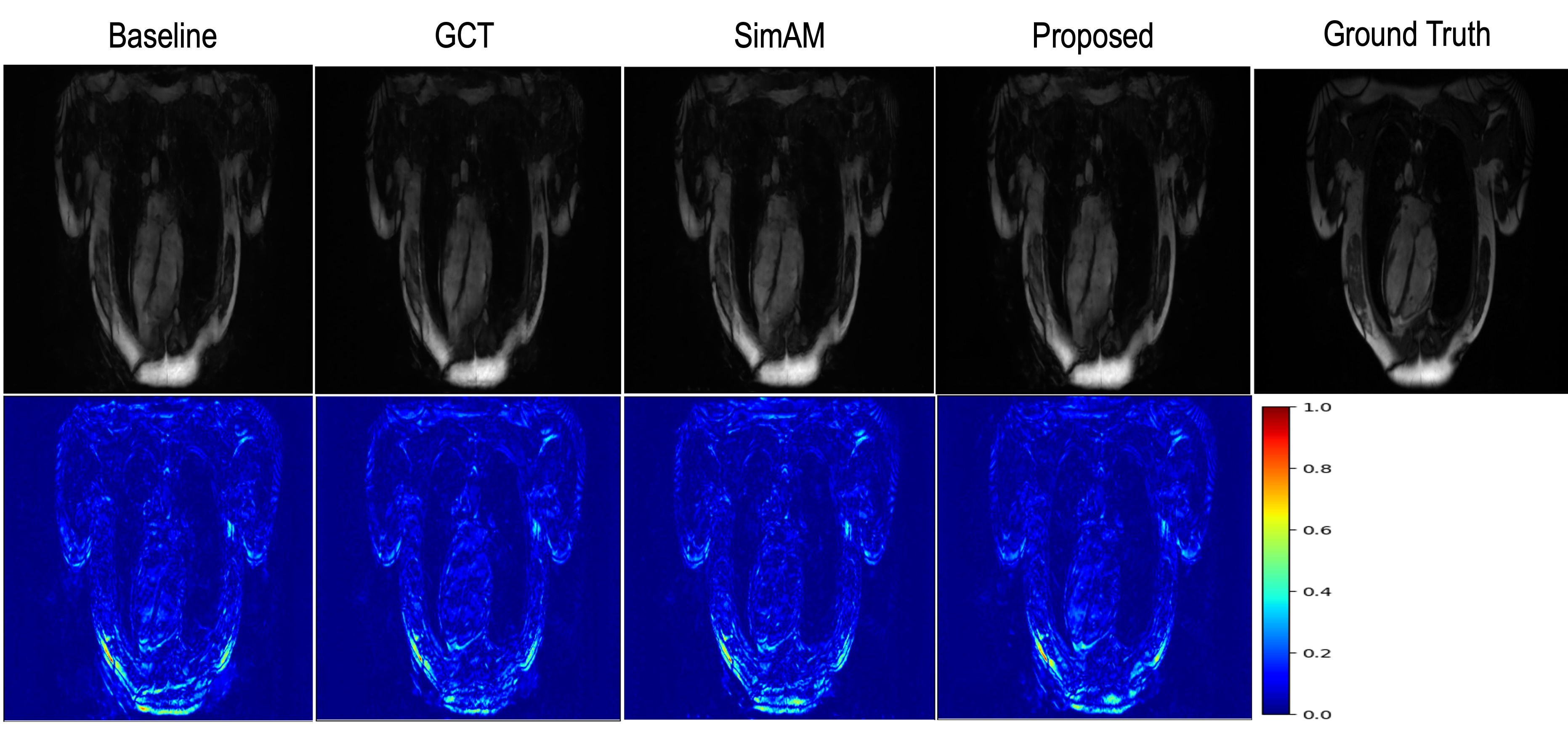}%{Q_plots_new.png}
%\vspace{-25pt}
\caption{The reconstruction results and normalized absolute error maps for a long-axis cine image (case P1) with a ×10 acceleration factor, across different methods.}
\vspace{-25pt}
\end{figure}
\section{Conclusion}
%In this study, we introduced a parameter-free attention mechanism inspired by ecological principles to enhance cardiac MRI reconstruction. 
Our approach, based on non-linear ecological difference equations drawn from established ecological principles, outperformed existing parameter-free methods underscoring its effectiveness in enhancing cardiac MRI reconstruction.

%We have verified that a nonlinear ecological mathematical model can be used to design parameter-free attention mechanism in the context of cardiac MRI reconstruction. In the future we plan to test our attention layer on classification, segmentation and interpretability tasks in medical imaging.

\begin{credits}
\subsubsection{\ackname} This publication has emanated from research conducted with the financial support of Science Foundation Ireland under Grant numbers 18/CRT/6183 and 12/RC/2289\_P2.

\subsubsection{\discintname}
The authors have no competing interests to declare that are
relevant to the content of this article. 
\end{credits}
%
% ---- Bibliography ----
%
% BibTeX users should specify bibliography style 'splncs04'.
% References will then be sorted and formatted in the correct style.
%
% \bibliographystyle{splncs04}
% \bibliography{mybibliography}
%

\end{document}